\documentclass[useAMS,usenatbib,usegraphicx]{mn2e}

\title[Vertical temperature structure of protoplanetary discs]
{Effects of scattering and dust grain size on the temperature structure
of protoplanetary discs: A three-layer approach}
\author[A. K. Inoue, A. Oka, \& T. Nakamoto]
{Akio K. Inoue$^{1}$\thanks{E-mail: akinoue@las.osaka-sandai.ac.jp}, 
Akinori Oka$^{2}$, and Taishi Nakamoto$^{2}$\\
$^{1}$College of General Education, Osaka Sangyo University, 3-1-1,
Nakagaito, Daito, Osaka 574-8530, Japan\\
$^{2}$Department of Earth and Planetary Sciences, Tokyo Institute of
Technology, Ookayama, Meguro, Tokyo 152-8551, Japan}
\begin{document}

\date{}

\pagerange{\pageref{firstpage}--\pageref{lastpage}} \pubyear{2008}

\maketitle

\label{firstpage}

\begin{abstract}
The temperature in the optically thick interior of protoplanetary discs
 is essential for the interpretation of millimeter observations of the
 discs, for the vertical structure of the discs, for models of the disc
 evolution and the planet formation, and for the chemistry in the discs.
 Since large icy grains have a large albedo even in the infrared, the
 effect of scattering of the diffuse radiation in the discs on the
 interior temperature should be examined. We have performed a series of
 numerical radiation transfer simulations including isotropic scattering
 by grains with various typical sizes for the diffuse radiation as well
 as for the incident stellar radiation. We also have developed an
 analytic model including isotropic scattering to understand the physics
 concealed in the numerical results. With the analytic model, we have
 shown that the standard two-layer approach is valid only for grey
 opacity (i.e. grain size $\ga10$ \micron) even without scattering. A
 three-layer interpretation is required for grain size $\la10$ \micron. 
 When the grain size is 0.1--10 \micron, the numerical simulations show
 that isotropic scattering reduces the temperature of the disc
 interior. This reduction is nicely explained by the analytic
 three-layer model as a result of the energy loss by scatterings of the
 incident stellar radiation and of the warm diffuse radiation in the
 disc atmosphere. For grain size $\ga10$ \micron\ (i.e. grey
 scattering), the numerical simulations show that isotropic scattering
 does not affect the interior temperature. This is nicely explained by
 the analytic two-layer model; the energy loss by scattering in the disc
 atmosphere is exactly offset by the ``green-house effect'' due to
 scattering of the cold diffuse radiation in the interior.
\end{abstract}


\section{Introduction}

Protoplanetary discs are planet formation sites. We observe the
electro-magnetic radiation from the discs to understand their physical
conditions, and then, to know the planet formation, especially, its
beginning. Property of the radiation from a disc is essentially
determined by the structure of the disc. The structure follows the
response of the disc to the radiation from the central star. 
To interpret the radiation from protoplanetary discs, therefore, we
should have a robust link between the radiation from the central star
and the disc structure, in particular, the temperature structure.

The temperature structure of the discs is also essential for chemical
reactions in the discs. The evolution of various molecules in the discs
and the exchange of these molecules between the gas phase and the solid
phase on grains will be discussed in detail with the ALMA in near
future \citep[e.g.,][]{nom08}. The condensation front of icy molecules,
so-called 'snow line', is also determined by the temperature structure
of the discs \citep[e.g.,][]{sas00,oka09}. The location of the snow line
is very important because it significantly enhances the amount of solid
materials to make planetary cores and affects the supply of water to
rocky planets. The temperature structure also affects the evolution of
the discs themselves. The accretion activity in the discs is supposed to
be driven by the magnetorotational instability \citep[e.g.,][]{san00}. 
This requires a certain degree of the ionization of disc materials which
depends on the structure of the discs.

A milestone in the research of the vertical temperature structure of
protoplanetary discs is the work by \cite{cg97} (hereafter CG97). They
proposed a two-layer model consisting of the super-heated layer directly
exposed by the stellar radiation and the interior warmed by the
super-heated layer. This model explained the shape of the spectral
energy distribution of the discs very well. The computational cheapness
of the analytic approach makes the CG97 model very useful to compare
with a large sample of the discs observed. Therefore, some attempts to
refine the simple model of CG97 were performed 
\citep{chi01,dul01,dul03a,raf06,gar07}. 

Despite a great success of the CG97 model, numerical simulations of the
radiation transfer in the discs showed a significant decrement of the
equatorial temperature relative to the prediction by the CG97 model
\citep{dul03a}. The internal energy loss by the radiation at a long
wavelength where the optical depth of the disc is relatively small is 
suggested as a cause of the decrement \citep{dul02,dul03a}. In the CG97
model and refined ones, the wavelength dependence of dust opacity was
taken into account in terms of mean opacity. \cite{dul02} argued the
importance of using full wavelength dependent opacity. However, we
propose an alternative approach in this paper: a three-layer model with
mean opacity which reproduces the temperature reduction quite well. In
addition, we show that the two-layer approximation in the CG97 model is
valid only when the dust opacity is ``grey'' which is expected if the
size of dust grains is larger than about 10 \micron.

Effect of scattering on the vertical temperature structure was not
considered in the literature very much. Scattering of the stellar
radiation was taken into account analytically by \cite{cal91} and
numerically by \cite{dul03} who showed that the temperature of the disc
interior is slightly reduced by the scattering. How about scattering of
the diffuse disc radiation? As the size of grains increases, the
scattering albedo increases. In particular, the albedo of large icy
grains is close to unity even for the infrared wavelength. This may
affect the disc structure significantly. Nevertheless, it has not been
examined so far.

This paper discusses the effect of scattering of the diffuse radiation
as well as that of the stellar radiation on the vertical temperature
structure of protoplanetary discs. Since the albedo depends on the grain
size, we examine the scattering effect as a function of the grain size.
Although there are 2-D/3-D numerical radiation transfer codes available
publicly, most of them have a serious difficulty in solving the radiation
equilibrium in very high optical depth ($\tau\sim10^6$) found in
protoplanetary discs \citep{pas04,ste06}. The RADICAL developed by
\cite{dul00} can solve such a problem without any difficulty thanks to a
variable Eddington tensor method. However, it treats scattering of only
the stellar radiation. We present a variable Eddington factor code with
both scatterings of the stellar radiation and of the diffuse radiation
but in a 1-D geometry. We also present an analytic model to interpret the
numerical results. This simple model would be very useful to understand
the physics determining the temperature structure of protoplanetary discs.

The rest of this paper consists of three sections; in section 2, we 
develop a numerical radiation transfer code taking into account both of
the scatterings but only for isotropic case and show the obtained
numerical solutions. In section 3, we construct an analytic model to
interpret the numerical solutions and discuss the physical mechanism
determining the temperature structure in protoplanetary discs. In the
final section, we summarise our findings.

\section{Numerical radiation transfer with scattering}

Our method is an extension of the variable Eddington factor method
developed by \cite{dul02}; we include isotropic scattering of the
diffuse radiation as well as that of the stellar radiation. 
A disc is divided into many annuli in which the transfer of the diffuse
radiation is treated one-dimensionally along the normal axis of each
annulus with neglecting the radiation energy transport among
annuli. This approximation, so-called 1+1D approximation, would be
reasonable in an optically thick disc, but not at the near of the disc
inner edge nor in a self-shadowing region \citep[e.g.,][]{dul01}. 
The radiation from the central star is separated from the diffuse
radiation and is treated with the so-called grazing angle recipe. 
In this paper, we only consider some single annulus cases in order to
feature the effect of the scattering on the temperature structure along
the normal axis of the annulus. Therefore, we assume a grazing angle 
$\alpha=0.05$ radian throughout of the paper. The density structure
along the normal axis of the annulus is solved to be consistent with the
obtained temperature structure assuming the hydrostatic equilibrium.
In Appendix A, we describe how to obtain the numerical solution of the
diffuse radiation transfer with isotropic scattering in each annulus in
detail.

\subsection{A simple dust model}

\begin{figure}
 \centering
 \includegraphics[width=7cm]{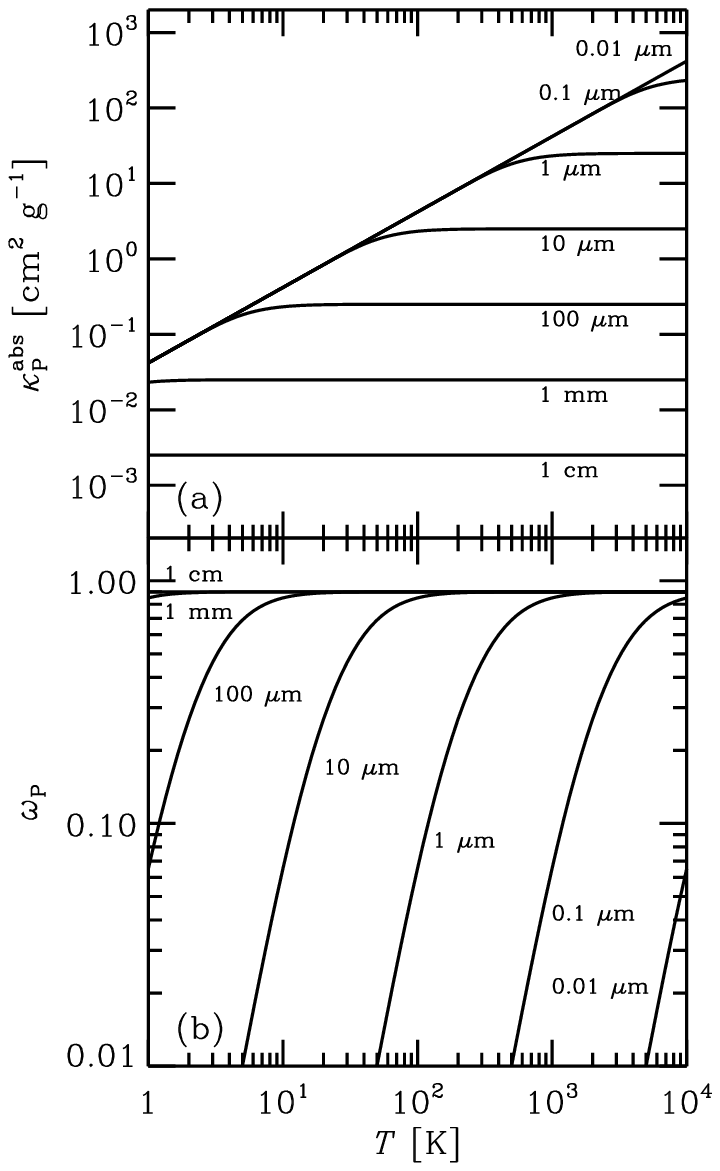}
 \caption{Planck mean properties of a simple dust model assumed in this
 paper: (a) absorption cross section per unit gas mass and (b) single
 scattering albedo for the case with $\omega_0=0.9$. We show seven cases
 of grain size from 0.01 \micron\ to 1 cm.}
\end{figure}

We adopt a very simple dust model in order to feature the effect of 
scattering on the temperature structure. The absorption and
scattering cross sections per unit gas mass at the wavelength $\lambda$
are assumed to be  
\begin{equation}
 \kappa_\lambda^{\rm abs} = 
  \cases{
  \kappa_0^{\rm abs} & ($\lambda \le \lambda_{\rm c}$) \cr
  \kappa_0^{\rm abs} \left(\frac{\lambda}{\lambda_{\rm c}}\right)^{-1}
  & ($\lambda > \lambda_{\rm c}$) \cr
  }\,,
\end{equation}
and 
\begin{equation}
 \kappa_\lambda^{\rm sca} = 
  \cases{
  \kappa_0^{\rm sca} & ($\lambda \le \lambda_{\rm c}$) \cr
  \kappa_0^{\rm sca} \left(\frac{\lambda}{\lambda_{\rm c}}\right)^{-4}
  & ($\lambda > \lambda_{\rm c}$) \cr
  }\,,
\end{equation}
respectively. The single scattering albedo at the wavelength $\lambda$
is 
\begin{equation}
 \omega_\lambda = \frac{\kappa_\lambda^{\rm sca}}
  {\kappa_\lambda^{\rm abs} + \kappa_\lambda^{\rm sca}}\,.
\end{equation}
The critical wavelength $\lambda_{\rm c}$ may be related to
a typical grain radius $a$ as $\lambda_{\rm c} = 2\pi a$.
If we consider a spherical grain composed of uniform material, the
absorption cross section is expressed as  
$\kappa_\lambda^{\rm abs}=(3{\cal D}Q_\lambda^{\rm abs})/(4\rho_{\rm d}a)$, 
where $Q_\lambda^{\rm abs}$ is the absorption cross section normalised
by the geometrical cross section $\pi a^2$, $\rho_{\rm d}$ is the
grain material density, and $\cal D$ is the dust-to-gas mass ratio.
With the values of ${\cal D}=10^{-2}$ (Solar system nebula), 
$\rho_{\rm d}=3$ g cm$^{-3}$ (silicate), and $Q_\lambda^{\rm abs} \to 1$
($\lambda \to 0$), we obtain the absorption cross section for small
wavelengths as 
\begin{equation}
 \kappa_0^{\rm abs} = 250~{\rm cm^{2}~g^{-1}}
  ~\left(\frac{\rm 0.1~\micron}{a}\right)\,.
\end{equation}
The scattering cross section can be given by the single scattering
albedo for small wavelengths:  
$\omega_0=\kappa_0^{\rm sca}/(\kappa_0^{\rm abs}+\kappa_0^{\rm sca})$.
In this paper, we consider three cases of $\omega_0=0$ (no scattering), 
$0.9$, or $0.99$. The values of $\omega_0$ for the last two cases may be
extreme but such a large albedo is expected for icy grains in some
wavelengths. 

Figure~1 shows Planck means of the absorption cross section and the
scattering albedo assumed in this paper as a function of the temperature
input into the Planck function. In the panels, we show seven cases of
grain size from 0.01 \micron\ to 1 cm. We note that the absorption cross
section and the scattering albedo become independent of the temperature,
i.e. ``grey'', when the temperature exceeds a critical one which depends
on the grain size, corresponds to the critical wavelength 
$\lambda_{\rm c}$, and is roughly expressed as  
$T_{\rm c}\sim10^3(1~\micron/a)$ K. 
In this paper, we do not consider the size distribution of the dust
grains. Thus, the ``grain size'' of this paper means a typical grain
size averaged over a size distribution function with a weight.

\subsection{Numerical results: Temperature structure}

We here show the results of the annulus with the radius of 1 AU obtained
from our numerical radiation transfer in Figures 2 and 3. The results
with other radii have been confirmed to be the same qualitatively. The
gas column density is assumed to be $10^3(R/{\rm AU})^{-1}$ g cm$^{-2}$,
where $R$ is the radial distance from the central star. The properties
of the central star assumed are the effective temperature $T_*=3,000$ K,
the radius $R_*=2.0$ $R_\odot$, and the mass $M_*=0.5$ $M_\odot$. Other
assumed parameters are as follows: the grazing angle $\alpha=0.05$, the
visible fraction of the stellar photosphere at the annuli 
$f_{\rm vis}=0.5$, and the mean molecular weight $\mu_{\rm m}=7/3$.

\begin{figure*}
 \centering
 \includegraphics[width=13cm]{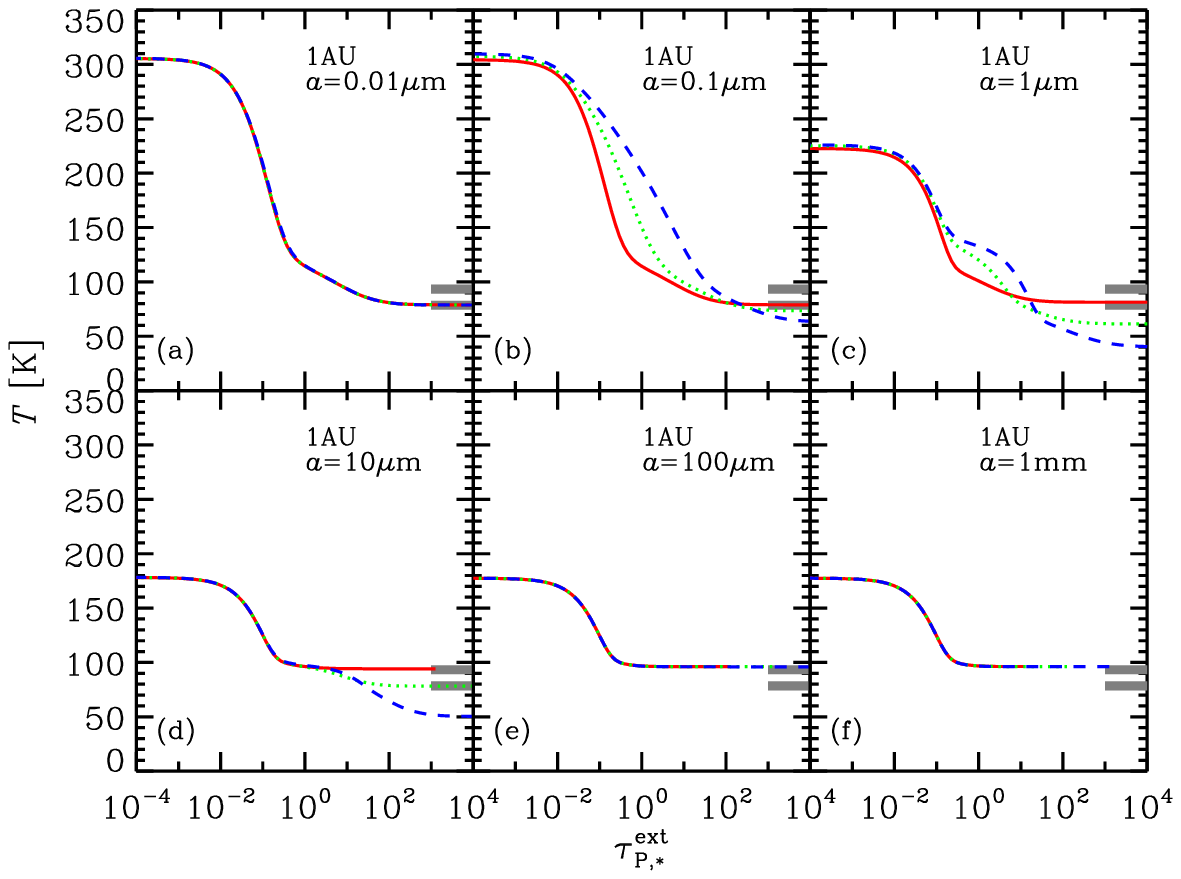}
 \caption{Vertical temperature structure in annuli with the radius of 1
 AU but with various grain sizes, $a$, shown in each panel. The
 horizontal axis is the Planck mean extinction optical depth with the
 stellar effective temperature ($T_*=3,000$ K). The solid lines are the
 no scattering case, the dotted lines are the case with the single
 scattering albedo for small wavelength $\omega_0=0.9$, and the dashed
 lines are the case with $\omega_0=0.99$. Two grey thick marks at the
 right-hand edge in each panel indicate the temperatures predicted by
 the analytic two-layer (upper mark; 93.3 K) or three-layer (lower mark;
 78.5 K) models without scattering presented in section 3.}
\end{figure*}

\begin{figure*}
 \centering
 \includegraphics[width=13cm]{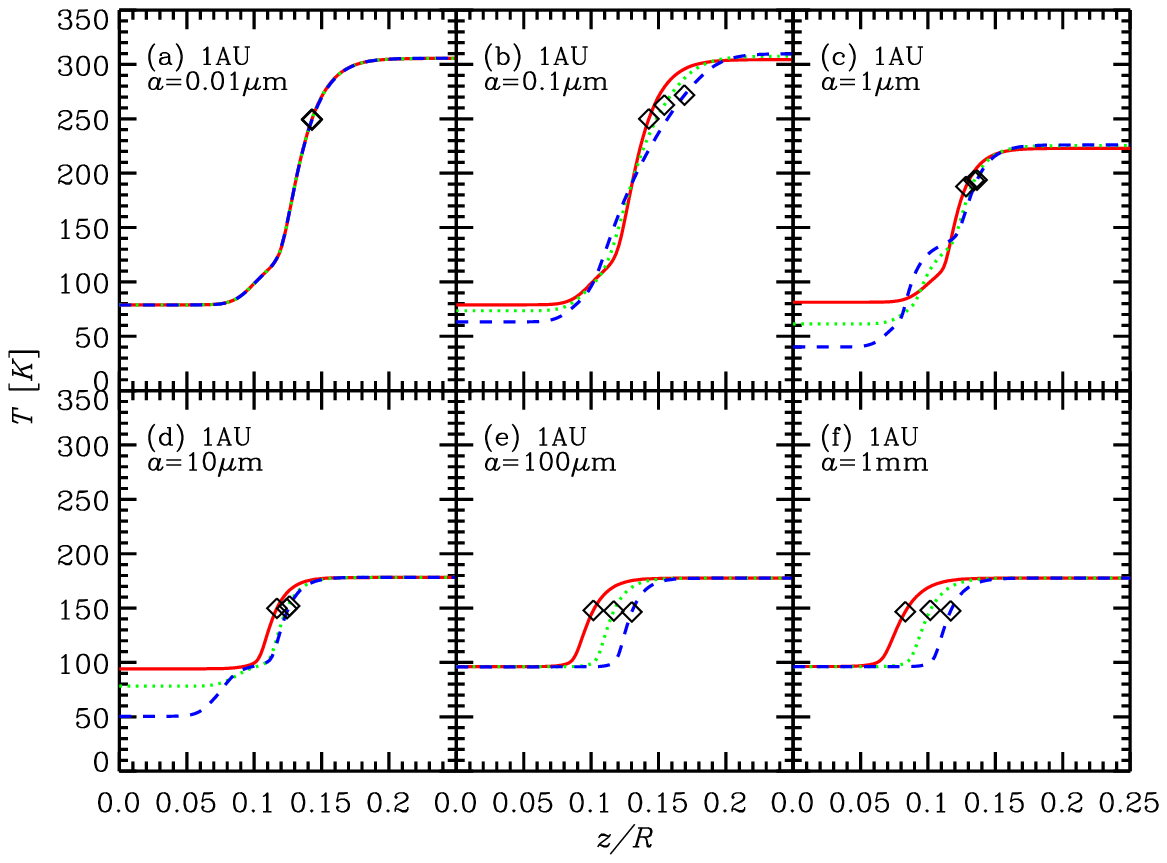}
 \caption{Same as Fig.~2, but the horizontal axis is the physical height
 from the equatorial plane normalised by the radius of the annulus. The
 diamonds indicate the lower boundary of the super-heated layer, at
 which the Planck mean extinction optical depth with the stellar
 effective temperature ($T_*=3,000$ K) is equal to the grazing angle
 ($\alpha=0.05$).}
\end{figure*}

Figure~2 shows the vertical temperature structures of annuli with 1 AU
radius with various grain sizes. We take a coordinate of the Planck mean
extinction optical depth with the stellar effective temperature as the
horizontal axis. Note that the maximum optical depth in each curve
occurs the equatorial plane. The grain sizes assumed are shown in each
panel. The solid, dotted, and dashed curves are the cases of no
scattering (i.e. $\omega_0=0$), $\omega_0=0.9$, and $\omega_0=0.99$,
respectively.

For no scattering cases (solid curves), the so-called two-layer
structure proposed by CG97 is confirmed. The dust temperature near the
surface is enhanced due to the direct stellar radiation: ``super-heated
layer''. The thickness of the super-heated layer is well expressed by
the Planck mean extinction optical depth as 
$\tau_{\rm P,*}^{\rm ext}\simeq\alpha=0.05$ (grazing angle). The
temperature rapidly decreases if $\tau_{\rm P,*}^{\rm ext}>\alpha$. If
the interior is optically thick against its own radiation, then, the
interior reaches the thermal equilibrium and becomes isothermal. As the
grain size becomes larger, the temperature of the super-heated layer
becomes lower. In contrast, the temperature of the interior becomes
higher. The physical reason of this phenomenon will be discussed in
section 3 with two analytic models: the standard two-layer model like
CG97 and a newly developed three-layer model. Here, we just mention the
fact that the numerical results agree with the prediction by the
three-layer model for the grain size of 0.01--1 \micron, whereas the
results agree with that by the two-layer model for the size $\ga10$
\micron.

When there is scattering, some differences appear. For $a=0.01$ \micron\
(panel [a]), the scattering albedo $\omega$ is negligible in the
wavelength interest (e.g., an effective temperature less than
$T_*=3,000$ K in Figure~1). Thus, scattering virtually has no effect. 
For $a=0.1$ \micron\ (panel [b]), $\omega$ for the stellar radiation is
significant, but that for the diffuse radiation in the annulus (its
effective temperature is less than about 300 K) is still negligible (see
Figure~1). In this case, the temperature at the equatorial plane becomes
slightly lower than that in the no scattering case, which is consistent
with \cite{dul03}. For $a=1$--10 \micron\ (panels [c,d]), $\omega$
becomes significant for the radiation of the super-heated layer. In
this case, we observe a plateau like structure at around 
$\tau_{\rm P,*}^{\rm ext}\sim1$ and a significant reduction of the
equatorial temperature. For $a\ga100$ \micron--1 mm (panels [e,f]),
finally, $\omega$ becomes ``grey'' for all the radiation considered
here. In this case, the temperature structure with scattering becomes
indistinguishable from that without scattering; a ``grey'' scattering
has no effect on the temperature structure in the optical depth
coordinate. The physical reasons of these features will be discussed in
section 3 with an analytic model.

Even when the ``grey'' scattering, we find a difference of the
temperature structures with/without scattering if we take the physical
height as the coordinate as shown in Figure~3. The diamonds in Figure~3
indicate the lower boundary of the super-heated layer, at which the
Planck mean extinction optical depth with the stellar effective
temperature ($T_*=3,000$ K) is equal to the grazing angle. We find that
with scattering, the height of the super-heated layer is always enhanced;
scattering causes more flaring disc \citep[see also][]{dul03}. This
suggests that the scattering may affect the global structure of the
disc, which will be discussed in a future work.

\section{Three-layer analytic model with scattering}

In order to understand the numerical results presented in the previous
section, we here develop an analytic model as an extension of the
seminal two-layer model by CG97: three-layer model with scattering. 
To describe fluxes across the boundaries of the layers, we adopt a
two-stream Eddington approximation with isotropic scattering. 
The notations in this section are summarised in Table~1.

\begin{table*}
 \caption{Notations in our analytic model.}
 \begin{tabular}{@{}lll}
  \hline
  Notation & Meaning & Remarks\\
  \hline
  $\alpha$ & Grazing angle of the entering stellar radiation &  \\
  $\Omega_*$ & Solid angle of the stellar photosphere & \\
  $W_*$ & Dilution factor of the stellar radiation & $\Omega_*/4\pi$ \\
  $H_*^{\rm in}$ & Stellar input flux & \\
  $H_x^{\rm up}$ & Upwards flux from the $x$ layer & \\
  $H_x^{\rm down}$ & Downwards flux from the $x$ layer & \\
  $H_{\rm input}$ & Total downwards flux from the super-heated layer & \\
  $T_x$ & Temperature of the $x$ layer & \\
  $B_x$ & Frequency integrated Planck function with $T_x$ & 
	  $(\sigma_{\rm SB}/\pi)T_x^4$ \\
  $\Sigma_x$ & Gas mass column density of the $x$ layer & \\
  $\kappa^{\rm abs}_x$ & Absorption cross section per unit gas mass for
      the radiation with $T_x$ & \\
  $\kappa^{\rm ext}_x$ & Extinction cross section per unit gas mass for
      the radiation with $T_x$ & \\
  $\tau^{\rm ext}_{x,y}$ & Extinction optical depth of the $x$ layer for
      the radiation with $T_y$ & 
	  $\kappa^{\rm ext}_y \Sigma_x$\\
  $\omega_x$ & Scattering albedo for the radiation with $T_x$ & \\
  $\chi_x$ & Square-root of the thermal coefficient $1-\omega_x$ & 
	  $\sqrt{1-\omega_x}$ \\
  $I^{\rm up}_{x,y}$ & Upwards intensity of the radiation with $T_y$
      from the $x$ layer & \\ 
  $I^{\rm down}_{x,y}$ & Downwards intensity of the radiation with $T_y$
      from the $x$ layer & \\ 
  $a_{x,y}$ & Thermal coefficient in the $x$ layer for the radiation
      with $T_y$ & eq.~(B13) \\
  $b_{x,y}$ & Reflection coefficient in the $x$ layer for the radiation
      with $T_y$ & eq.~(B14) \\
  $c_{x,y}$ & Transmission coefficient in the $x$ layer for the radiation
      with $T_y$ & eq.~(B15) \\
  $\Phi_{\rm input}$ & Ratio of $H_{\rm input}$ to $H_*^{\rm in}$ & 
	  eq.~(20) \\
  $\Phi_{\rm i(2)}$ & Reduction factor of $B_{\rm i}$ by scattering in
      the two-layer model & eq.~(23) \\
  $\Phi_{\rm i(3)}$ & Reduction factor of $B_{\rm i}$ by scattering in
      the three-layer model & eq.~(29) \\
  \hline
 \end{tabular}

 \medskip
 Subscripts $x$ and $y$ are * for stellar quantities, s for super-heated
 layer quantities, m for middle layer quantities, or i for interior
 quantities.\\
 $\sigma_{\rm SB}$ is the Stefan-Boltzmann constant.
\end{table*}

\subsection{Model description}

\begin{figure*}
 \centering
 \includegraphics[width=13cm]{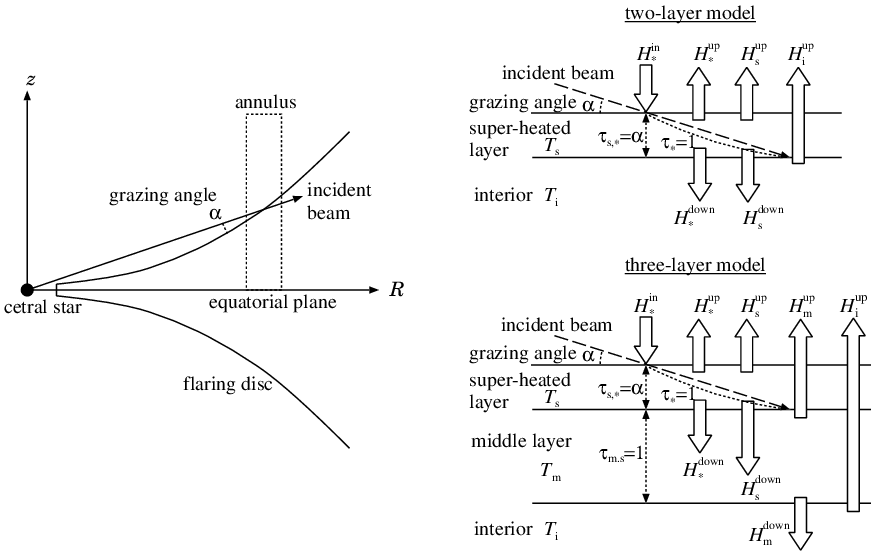}
 \caption{Schematic diagram of the two-layer and the three-layer
 models. We are considering an annulus clipped from a flaring disc
 surrounding a young star as shown in the left-hand picture.  In the
 annulus, we consider two or three layers as shown in the right-hand
 pictures. Each layer is assumed to be isothermal with a temperature
 $T$. The fluxes crossing the boundaries of layers are denoted as 
 $H^{\rm up}$ or $H^{\rm down}$ depending on their direction . We set
 the direction of arrows in the right-hand pictures as the positive
 direction for the fluxes. The subscript of each quantity indicates the
 layer associated with the quantity: ``s'' for the super-heated layer,
 ``m'' for the middle layer, and ``i'' for the interior. The quantities
 with the subscript of ``*'' are associated with the radiation from the
 central star and $H_*^{\rm in}$ is the stellar flux at the top of the
 annulus. See also Table~1 for notations.}
\end{figure*}

Suppose two or three layers in an annulus as shown in Figure~4. 
We assume that each layer is isothermal with the temperature determined
by the radiation equilibrium in the layer. Then, we consider
that each layer emits the radiation characterised by its temperature and
other layers just work as absorption and scattering media for the
radiation. The radiation from the central star is characterised by the
stellar effective temperature. The characterisation of the radiation is
done by the Planck mean and the characteristic frequencies are denoted
by each subscript such as ``*'' for the stellar radiation (see the
caption of Figure~4 and Table~1). We denote, for example, the extinction
cross section characterised by the stellar effective temperature as 
$\kappa_*^{\rm ext}$. Note that all the quantities depending on the
frequency are Planck averaged in this section.

We call the two or three layers super-heated layer, middle layer, and
interior as shown in Figure~4. The super-heated layer is defined as only
the layer exposed by the direct stellar radiation entering into the
annulus with a small grazing angle $\alpha$. The optical thickness of
the layer is $\approx\alpha$ as shown by the numerical solutions in
\S2.2. Thus, we define the thickness of the super-heated layer as 
$\tau_{\rm s,*}^{\rm ext}\equiv\kappa_*^{\rm ext}\Sigma_{\rm s}=\alpha$, 
where $\Sigma_{\rm s}$ is the gas column density of the layer. 
The interior represents the isothermal part found in the numerical
solutions. Thus, the boundary can be defined by the photosphere of its
own radiation. However, we here simply define the interior as the part
other than the super-heated and the middle layers.

The middle layer is introduced by the following consideration. When the
opacity coefficient decreases as the wavelength increases, the
absorption of the radiation from the warm super-heated layer occurs well
above the photosphere of the cold interior radiation. In this case, the
interior is not warmed directly by the super-heated layer but by the
``middle'' layer where the radiation of the super-heated layer is
effectively absorbed. We here define the thickness of the middle layer as 
$\tau_{\rm m,s}^{\rm ext}\equiv\kappa_{\rm s}^{\rm ext}\Sigma_{\rm m}=1$
with the gas column density of the middle layer $\Sigma_{\rm m}$ 
although this definition is rather arbitrary. On the other hand, when
the opacity coefficient is grey, the middle layer with the above
thickness is optically thick for its own radiation. Thus, the middle
layer reaches the thermal equilibrium and is merged into the isothermal
interior. In this case, we do not need to consider the middle layer. 
Therefore, we have two cases: the three-layer model when 
$\tau_{\rm m,m}^{\rm ext}\equiv\kappa_{\rm m}^{\rm ext}\Sigma_{\rm m}<1$
and the two-layer model when $\tau_{\rm m,m}^{\rm ext}\simeq1$ (or 
$\tau_{\rm m,i}^{\rm ext}\equiv\kappa_{\rm i}^{\rm ext}\Sigma_{\rm m}\simeq1$ 
because the middle layer is merged into the interior). 
In other words, three layers are needed when 
$\kappa_{\rm m}^{\rm ext}/\kappa_{\rm s}^{\rm ext}<1$
and two layers are sufficient when 
$\kappa_{\rm i}^{\rm ext}/\kappa_{\rm s}^{\rm ext}\simeq1$.
Tables~2 and 3 are summaries of the thickness of the upper two layers and
the condition of the two or three-layer models.
Importantly, the seminal two-layer model is valid only when the opacity
coefficient is grey in the frequencies interest. This fact has not
seemed to be known well so far.

\begin{table}
 \caption{Thickness of the upper two layers.}
 \begin{tabular}{@{}ll}
  \hline
  Super-heated layer & $\tau^{\rm ext}_{\rm s,*}=\alpha$, i.e.  
  $\Sigma_{\rm s}=\alpha/\kappa^{\rm ext}_*$\\
  Middle layer & $\tau^{\rm ext}_{\rm m,s}=1$, i.e.
  $\Sigma_{\rm m}=1/\kappa^{\rm ext}_{\rm s}$\\
  \hline
 \end{tabular}
\end{table}

\begin{table}
 \caption{Condition of the two or three-layer models.}
 \begin{tabular}{@{}ll}
  \hline
  Three-layer model & 
  $\kappa^{\rm ext}_{\rm m}/\kappa^{\rm ext}_{\rm s}<1$\\
  Two-layer model & 
  $\kappa^{\rm ext}_{\rm i}/\kappa^{\rm ext}_{\rm s}\simeq1$\\
  \hline
 \end{tabular}
\end{table}

\subsubsection{Stellar fluxes}

When the grazing angle $\alpha$ is small, the stellar flux at the top of
the annulus is 
\begin{equation}
 H_*^{\rm in} = \alpha W_* B_*\,,
\end{equation}
with the integrated Planck function $B_*=(\sigma_{\rm SB}/\pi)T_*^4$ 
and the dilution factor $W_*=\Omega_*/4\pi$, where $\Omega_*$ is the
solid angle of the stellar photosphere from the top of the annulus. 
If only the fraction $f_{\rm vis}$ of the stellar photosphere is
visible because of the optically thick disc, the solid angle becomes 
$\Omega_*=f_{\rm vis}\pi(R_*/R)^2$, where $R_*$ is the stellar radius
and $R$ is the radius of the annulus.

When there is scattering, a part of the incident stellar flux is
reflected upwards and downwards by the super-heated layer. \cite{cal91}
presented an analytic expression of the scattered flux for isotropic
scattering. From equation (5) in \cite{cal91}, the outbound fluxes
at the upper and lower boundaries of the super-heated layer (see
Figure~4) become 
\begin{equation}
 H_*^{\rm up} = \alpha W_* B_* 
  \left[\frac{\omega_*}{1+\chi_*}\right]\,,
\end{equation}
and 
\begin{equation}
 H_*^{\rm down} = \alpha W_* B_* 
  \left[\frac{\omega_* \chi_*}{1+\chi_*}\right]\,,
\end{equation}
where $\omega_*$ is the single scattering albedo at the stellar
frequency and $\chi_*=\sqrt{1-\omega_*}$. In the derivation of
equations (6) and (7), we have assumed 
$\tau_{\rm s,*}^{\rm ext}\equiv\kappa_*^{\rm ext}\Sigma_{\rm s}=\alpha\ll1$, 
adopted a different upper boundary condition from
\cite{cal91}\footnote{\cite{cal91} adopted $J(0)=2H(0)$, where $J(0)$
and $H(0)$ are the mean intensity and the mean flux of the scattered
stellar radiation at the top of the medium. On the other hand, we
adopted $J(0)=\sqrt{3}H(0)$. This difference is due to the choice of the
angle of the stream line.}, and neglected the term $e^{-1}$ for the
downwards flux. Note that the total scattered flux is 
$H^{\rm up}_* + H^{\rm down}_* 
= \omega_* \alpha W_* B_* = \omega_* H^{\rm in}_*$ and 
$H_*^{\rm up}=H_*^{\rm down}=0$ if $\omega_*=0$ (no scattering case).

\subsubsection{Super-heated layer fluxes}

The radiation characterised by the temperature of the super-heated layer
$T_{\rm s}$ is produced only in the super-heated layer. In other layers,
this radiation is not produced but is absorbed or scattered. The
super-heated layer is vertically optically thin for its own radiation as 
$\tau_{\rm s,s}^{\rm ext}\equiv\kappa_{\rm s}^{\rm ext}\Sigma_{\rm s}
=(\kappa_{\rm s}^{\rm ext}/\kappa_*^{\rm ext})\alpha\ll1$ because the
grazing angle $\alpha$ is small and we have  
$(\kappa_{\rm s}^{\rm ext}/\kappa_*^{\rm ext})\le1$. On the other hand, 
the total optical depth of other layers is very large. Thus, we consider
a geometry that a thin isothermal layer lies on a semi-infinite
absorption and scattering slab. 

The above of the super-heated layer is assumed to be vacuum; there is no
downwards input radiation with the temperature $T_{\rm s}$ at the top of
the layer. However, there is upwards input radiation at the bottom of
the layer because of the reflection by the semi-infinite interior below
the layer. The upwards and downwards radiation intensities from the
super-heated layer ($I^{\rm up}_{\rm s,s}$ and $I^{\rm down}_{\rm s,s}$,
respectively) in the two-stream Eddington approximation (the cosine of
the angle between the stream lines and the normal of the layer is set to
be $\pm1/\sqrt{3}$) become from equations (B12) and (B16) 
\begin{equation}
 I^{\rm up}_{\rm s,s} = a_{\rm s,s} B_{\rm s} 
  + c_{\rm s,s} I^{\rm up}_{\rm i,s}\,,
\end{equation}
and 
\begin{equation}
 I^{\rm down}_{\rm s,s} = a_{\rm s,s} B_{\rm s} 
  + b_{\rm s,s} I^{\rm up}_{\rm i,s}\,, 
\end{equation}
where $I^{\rm up}_{\rm i,s}$ is the upwards input intensity from the
interior, and $a_{\rm s,s}$, $b_{\rm s,s}$, and $c_{\rm s,s}$ are the
thermal, reflection, and transmission coefficients in the super-heated
layer for the radiation characterised by the temperature $T_{\rm s}$
(see eqs.~[B13--B15]). The reflected intensity from the interior becomes 
\begin{equation}
 I^{\rm up}_{\rm i,s}=b_{\rm i,s}I^{\rm down}_{\rm s,s}\,,
\end{equation}
where $b_{\rm i,s}$ is the reflection coefficient in the interior for
the radiation with $T_{\rm s}$. Note that the interior itself does not
emit radiation with $T_{\rm s}$.

We can always obtain the unique exact solution of $I^{\rm up}_{\rm s,s}$, 
$I^{\rm down}_{\rm s,s}$, and $I^{\rm up}_{\rm i,s}$ from equations
(8--10), whereas we derive an approximate solution of them in this paper.
As found equation (B18), for a semi-infinte slab, we have 
$b_{\rm i,s}\approx(1-\chi_{\rm s})/(1+\chi_{\rm s})$,
where $\chi_{\rm s}=\sqrt{1-\omega_{\rm s}}$ with $\omega_{\rm s}$ being
the scattering albedo for the radiation with $T_{\rm s}$. Since the
super-heated layer is optically thin for its own radiation, we can
approximate $b_{\rm s,s}\ll1$ and $c_{\rm s,s}\approx1$. We also find
$a_{\rm s,s}\approx\sqrt{3}\chi_{\rm s}^2\tau_{\rm s,s}^{\rm ext}
=\sqrt{3}\chi_{\rm s}^2(\kappa_{\rm s}^{\rm ext}/\kappa_*^{\rm ext})\alpha$ 
from equation (B13) for a small optical depth. Then, we obtain 
$I^{\rm up}_{\rm s,s}\approx (1+b_{\rm i,s})a_{\rm s,s}B_{\rm s}$, 
$I^{\rm down}_{\rm s,s}\approx a_{\rm s,s}B_{\rm s}$, 
and $I^{\rm up}_{\rm i,s}\approx b_{\rm i,s}a_{\rm s,s}B_{\rm s}$.
The upwards and downwards fluxes from the super-heated layer are 
$H^{\rm up}_{\rm s}=I^{\rm up}_{\rm s,s}/(2\sqrt{3})$ and 
$H^{\rm down}_{\rm s}=(I^{\rm down}_{\rm s,s}-I^{\rm up}_{\rm i,s})/(2\sqrt{3})$.
Therefore, we obtain 
\begin{equation}
 H_{\rm s}^{\rm up} = \alpha B_{\rm s}
  \left(\frac{\kappa_{\rm s}^{\rm ext}}{\kappa_*^{\rm ext}}\right)
  \left[\frac{\chi_{\rm s}^2}{1+\chi_{\rm s}}\right]\,,
\end{equation}
and 
\begin{equation}
 H_{\rm s}^{\rm down} = \alpha B_{\rm s}
  \left(\frac{\kappa_{\rm s}^{\rm ext}}{\kappa_*^{\rm ext}}\right)
  \left[\frac{\chi_{\rm s}^3}{1+\chi_{\rm s}}\right]\,.
\end{equation}

\subsubsection{Middle layer fluxes}

As discussed in the section 3.1, we consider the middle layer only when
the layer is optically thin for its own radiation. Although the middle
layer is sandwiched between the super-heated layer and the interior, we
neglect the effect of the super-heated layer because the optical depth
of the layer is very small (see above). The optical depth of the
interior is so large that we can regard it as a semi-infinite
medium. Thus, we have the same setting as the super-heated layer, other
than the optical thickness of the layer, $\tau_{\rm m,m}^{\rm ext}
=(\kappa_{\rm m}^{\rm ext}/\kappa_{\rm s}^{\rm ext})$.
Following the section 3.1.2, we have 
\begin{equation}
 H_{\rm m}^{\rm up} = B_{\rm m}
  \left(\frac{\kappa_{\rm m}^{\rm ext}}{\kappa_{\rm s}^{\rm ext}}\right)
  \left[\frac{\chi_{\rm m}^2}{1+\chi_{\rm m}}\right]\,, 
\end{equation}
and 
\begin{equation}
 H_{\rm m}^{\rm down} = B_{\rm m}
  \left(\frac{\kappa_{\rm m}^{\rm ext}}{\kappa_{\rm s}^{\rm ext}}\right)
  \left[\frac{\chi_{\rm m}^3}{1+\chi_{\rm m}}\right]\,, 
\end{equation}
where $B_{\rm m}$ is the integrated Planck function with the temperature
of the middle layer $T_{\rm m}$, $\omega_{\rm m}$ is the scattering
albedo for the radiation of the middle layer, and 
$\chi_{\rm m}=\sqrt{1-\omega_{\rm m}}$.

\subsubsection{Interior flux}

We always consider the interior to be optically thick for its own
radiation. On the other hand, other layers above the interior are
considered to be always optically thin. If we neglect the effect of
the upper layers, the interior is regarded as a semi-infinite isothermal
medium without incident flux of the radiation with the temperature
$T_{\rm i}$. In this case, based on equation (B12), the
upwards outbound flux becomes 
$H_{\rm i}^{\rm up} = a_{\rm i,i} B_{\rm i} / (2\sqrt{3})$ with 
$a_{\rm i,i}$ being the thermal coefficient in the interior for the
radiation with $T_{\rm i}$ and $B_{\rm i}$ being the integrated Planck
function with $T_{\rm i}$. For a semi-infinite medium, 
$a_{\rm i,i}\approx2\chi_{\rm i}/(1+\chi_{\rm i})$, 
where $\chi_{\rm i}=\sqrt{1-\omega_{\rm i}}$ with
$\omega_{\rm i}$ being the single scattering albedo for the interior
radiation (eq.~[B17]). Therefore, we have  
\begin{equation}
 H_{\rm i}^{\rm up}=\frac{B_{\rm i}}{\sqrt{3}}
  \left[\frac{\chi_{\rm i}}{1+\chi_{\rm i}}\right]\,.
\end{equation}
Note that the factor in $[~~~]$ is equal or less than 1/2. The equal is
true without scattering. Therefore, in general, scattering reduces the
radiation energy loss from the interior, i.e. the green-house effect.

\subsection{Temperature of the super-heated layer}

The radiation energy conservation in the super-heated layer is 
\begin{equation}
 H_*^{\rm in} - H_*^{\rm up} - H_*^{\rm down} 
  = H_{\rm s}^{\rm up} + H_{\rm s}^{\rm down}\,.
\end{equation}
The left-hand side is the net input flux of the stellar radiation and
becomes $\alpha W_* B_* (1-\omega_*)$. When there is scattering, the
input flux is reduced by a factor of $1-\omega_*$. The right-hand side is
the net output flux of the super-heated layer and becomes 
$\alpha B_{\rm s}(\kappa_{\rm s}^{\rm ext}/\kappa_*^{\rm ext})
(1-\omega_{\rm s})$. Thus we obtain 
\begin{equation}
 B_{\rm s} = B_* W_* 
  \left(\frac{\kappa_*^{\rm abs}}{\kappa_{\rm s}^{\rm abs}}\right)\,,
\end{equation}
or 
\begin{equation}
 T_{\rm s} = T_* W_*^{1/4} 
  \left(\frac{\kappa_*^{\rm abs}}{\kappa_{\rm s}^{\rm abs}}\right)^{1/4}\,,
\end{equation}
where we have applied $\kappa^{\rm abs}=(1-\omega) \kappa^{\rm ext}$.
Therefore, we expect $T_{\rm s}$ to be independent of the existence of
scattering. This result can be recovered by a microscopic consideration
for each dust grain in the radiation equilibrium with the stellar
radiation field in the super-heated layer. We also expect that 
$T_{\rm s}$ shows two asymptotic values since 
$(\kappa_*^{\rm abs}/\kappa_{\rm s}^{\rm abs})=T_*/T_{\rm s}$ 
for a small grain size and 
$(\kappa_*^{\rm abs}/\kappa_{\rm s}^{\rm abs})={\rm constant}$ for 
a large grain size as shown in Figure~1.

\begin{figure}
 \centering
 \includegraphics[width=7cm]{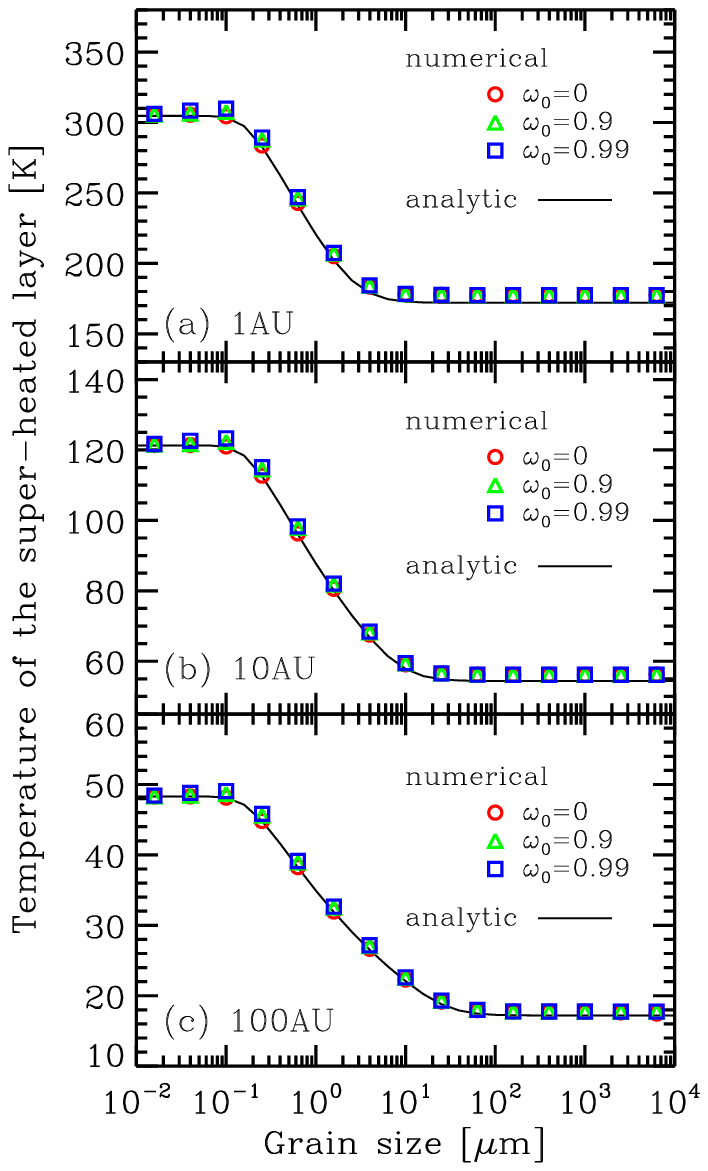}
 \caption{Temperature of the super-heated layer as a function of grain
 size: (a) annulus with the radius of 1 AU, (b) 10 AU, and (c) 100 AU. 
 The numerical solutions are shown by symbols; circles are no
 scattering case, triangles are the case with the single scattering
 albedo for small wavelength $\omega_0=0.9$, and squares are the case
 with $\omega_0=0.99$. The solid lines are the analytic model given by
 equation (18).}
\end{figure}

Figure~5 shows the temperature of the super-heated layer as a function
of grain size. The numerical solutions presented in section 2 are
shown by symbols: circles for the no scattering case, triangles for 
the $\omega_0=0.9$ case, and squares for the
$\omega_0=0.99$ case. The analytic model presented in equation (18) is
shown by the solid lines. To obtain the solution $T_{\rm s}$ of equation
(18), we need an iterative procedure because $\kappa^{\rm abs}_{\rm s}$
depends on $T_{\rm s}$. Figure~5 first shows that the scattering hardly
affect $T_{\rm s}$ as expected by the analytic model. Indeed, the cases
of the three different albedos are superposed for almost all grain
sizes, although a temperature enhancement of a few percent is observed
in the scattering cases at a grain size of 0.1--1 \micron. Figure~5 
second shows that the numerical solutions can be divided into two cases:
higher temperature for smaller grain size ($\la 0.1$ \micron) and lower
temperature for larger grain size ($\ga 10$ \micron), which is also
expected by the analytic model. The agreement between the numerical
solutions and the analytic model is excellent although the numerical
solutions give about 3\% higher temperature than the analytic model. 
This small difference is probably caused by neglecting the absorption of
the interior radiation in the super-heated layer in the analytic model.

\subsection{Temperature of the interior}

We can define the radiation flux input into the interior (including the
middle layer) of the annulus as 
\begin{equation}
 H_{\rm input} \equiv H_*^{\rm down} + H_{\rm s}^{\rm down} 
  = \alpha W_* B_* \Phi_{\rm input}\,,
\end{equation}
where 
\begin{equation}
 \Phi_{\rm input} = \frac{\omega_*\chi_*}{1+\chi_*}
  + \frac{\chi_*^2 \chi_{\rm s}}{1+\chi_{\rm s}}\,, 
\end{equation}
and we have eliminated $B_{\rm s}$ by equation (17). If there is no
scattering ($\omega_*=\omega_{\rm s}=0$, i.e. $\chi_*=\chi_{\rm s}=1$), 
we have $\Phi_{\rm input}=1/2$. That is, 
$H_{\rm input}=\alpha W_* B_*/2=H_*^{\rm in}/2$; a half of
the radiation energy received at the top of the annulus is input into
the interior \citep{cg97}.  For the isotropic scattering case, we always
have $H_{\rm input}<H_*^{\rm in}/2$; isotropic scattering always reduces
the energy input into the interior relative to the no scattering case 
\citep{dul03}.

\subsubsection{Two-layer model}

The energy conservation in the interior under the two-layer model is 
\begin{equation}
 H_{\rm input} = H_{\rm i}^{\rm up}\,.
\end{equation}
From equations (15) and (19), we obtain 
\begin{equation}
 B_{\rm i} = \sqrt{3} \alpha W_* B_* \Phi_{\rm i(2)}\,,
\end{equation}
where 
\begin{equation}
 \Phi_{\rm i(2)} = \left(\frac{1+\chi_{\rm i}}{\chi_{\rm i}}\right) 
  \Phi_{\rm input}\,.
\end{equation}
The interior temperature becomes 
\begin{equation}
 T_{\rm i} = T_* \left[\sqrt{3}\alpha W_* \Phi_{\rm i(2)}\right]^{1/4}\,.
\end{equation}
Therefore, $T_{\rm i}$ is proportional to a factor of 
$\Phi_{\rm i(2)}^{1/4}$. When there is no scattering 
($\Phi_{\rm input}=1/2$ and $\chi_{\rm i}=1$), we have 
$\Phi_{\rm i(2)}=1$.

As discussed in section 3.1, the two-layer model is valid if the
opacity coefficient is regarded as grey at all the frequencies interest: 
$\kappa_{\rm i}^{\rm ext}/\kappa_{\rm s}^{\rm ext}\simeq1$.
If this condition is satisfied, the scattering is also grey; 
$\omega_*=\omega_{\rm s}=\omega_{\rm i}$ (i.e. 
$\chi_*=\chi_{\rm s}=\chi_{\rm i}$). In this case, we have 
\begin{equation}
 \Phi_{\rm i(2)} = 1~~~{\rm (no/grey~scattering)}\,.
\end{equation}
Importantly, the factor $\Phi_{\rm i(2)}$ for grey isotropic scattering
is independent of the albedo and equal to the case without scattering. 
This is caused by the fact that the reduction of the flux input into the
interior by scattering ($\Phi_{\rm input}$ in eq.[19]) is completely
offset by the reduction of the flux outbound from the interior by
scattering ($(1+\chi_{\rm i})/\chi_{\rm i}$ in eq.[15]) for grey and
isotropic scattering. Therefore, we expect the same interior temperature
for grey and isotropic scattering case as that for no scattering case in
the two-layer model.

\subsubsection{Three-layer model}

The energy conservations in the middle layer and the interior under the
three-layer model are 
\begin{equation}
 H_{\rm input} = H_{\rm m}^{\rm up} + H_{\rm m}^{\rm down}\,,
\end{equation}
and 
\begin{equation}
 H_{\rm m}^{\rm down} = H_{\rm i}^{\rm up}\,.
\end{equation}
Since $H_{\rm m}^{\rm down} = \chi_{\rm m} H_{\rm m}^{\rm up}$ from
equations (13) and (14), we obtain 
$H_{\rm i}^{\rm up}=\chi_{\rm m}/(1+\chi_{\rm m}) H_{\rm input}$. Thus, 
\begin{equation}
 B_{\rm i} = \sqrt{3} \alpha W_* B_* \Phi_{\rm i(3)}\,,
\end{equation}
where 
\begin{equation}
 \Phi_{\rm i(3)} = \left(\frac{\chi_{\rm m}}{\chi_{\rm i}}\right)
   \left(\frac{1+\chi_{\rm i}}{1+\chi_{\rm m}}\right) \Phi_{\rm input}\,.
\end{equation}
The interior temperature becomes 
\begin{equation}
 T_{\rm i} = T_* \left[\sqrt{3}\alpha W_* \Phi_{\rm i(3)}\right]^{1/4}\,.
\end{equation}
Therefore, $T_{\rm i}$ is proportional to a factor of 
$\Phi_{\rm i(3)}^{1/4}$.

From the discussion in section 3.1, the three layer model is valid when 
$\kappa_{\rm m}^{\rm ext}/\kappa_{\rm s}^{\rm ext}<1$. This condition is
satisfied when the grain size, $a$, is smaller than 10--100 \micron.
When $a\la0.01$ \micron, the scattering albedo is
negligible at the all frequencies interest. In this case or the no
scattering case ($\omega_x=0$, i.e. $\chi_x=1$), we have  
\begin{equation}
 \Phi_{\rm i(3)} = \frac{1}{2}~~~{\rm (no~scattering)}\,.
\end{equation}
Comparing this with the two-layer model, we find that $T_{\rm i}$ in the 
three-layer model is reduced by a factor of $(1/2)^{1/4}$ relative to
that in the two-layer model even for no scattering case.

When $0.01\la a \la0.1$ \micron, 
$\omega_{\rm i}\approx\omega_{\rm m}\approx\omega_{\rm s}\approx0$ 
($\chi_{\rm i}\approx\chi_{\rm m}\approx\chi_{\rm s}\approx1$) 
but $\omega_*>0$ ($\chi_*<1$). In this case, we have 
\begin{equation}
 \Phi_{\rm i(3)} = \frac{\chi_*}{2}(2-\chi_*)\,.
\end{equation}
Thus, we find $\Phi_{\rm i(3)}<1/2$; we expect a reduction of $T_{\rm
i}$ relative to that without scattering.
When $0.1\la a \la1$--10 \micron, 
$\omega_{\rm i}\approx\omega_{\rm m}\approx0$ 
($\chi_{\rm i}\approx\chi_{\rm m}\approx1$) and 
$\omega_{\rm s}\approx\omega_*>0$ ($\chi_{\rm s}\approx\chi_*<1$). 
In this case, we have 
\begin{equation}
 \Phi_{\rm i(3)} = \frac{\chi_*}{1+\chi_*}\,.
\end{equation}
When 1--$10\la a \la10$--100 \micron, 
$\omega_{\rm i}\approx0$ ($\chi_{\rm i}\approx1$) and 
$\omega_{\rm m}\approx\omega_{\rm s}\approx\omega_*>0$ 
($\chi_{\rm m}\approx\chi_{\rm s}\approx\chi_*<1$). Then, we have 
\begin{equation}
 \Phi_{\rm i(3)} = 2\left(\frac{\chi_*}{1+\chi_*}\right)^2\,.
\end{equation}
When $a\ga10$--100 \micron, the opacity becomes almost grey, thus, the
three-layer model is no longer valid. We should choose the two-layer
model in this case.

\begin{figure}
 \centering
 \includegraphics[width=7cm]{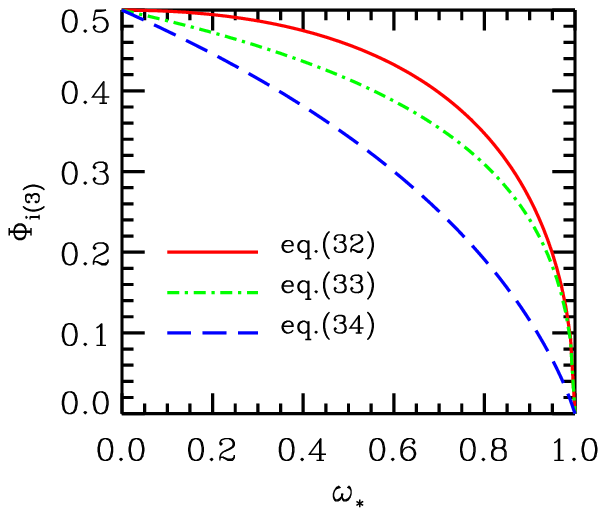}
 \caption{Reduction factor of the interior flux by isotropic scattering
 in the analytic three-layer model, $\Phi_{\rm i,(3)}$, as a function of
 the albedo for the stellar radiation, $\omega_*$. The solid, dot-dashed,
 and dashed lines indicate the cases with a typical grain size of 
 $0.01\la a\la0.1$ \micron, $0.1\la a\la1$--10 \micron, and 
 1--$10\la a\la10$--100 \micron, respectively. The reduction of the
 interior temperature relative to that in the two-layer model without
 scattering is given by $\Phi_{\rm i,(3)}^{1/4}$. Note that the interior
 temperature in the three-layer model is reduced by a factor of
 $(1/2)^{1/4}$ even for $\omega_*=0$ (i.e. no scattering).}
\end{figure}

Figure~6 shows the scattering reduction factor $\Phi_{\rm i,(3)}$ given
by equations (32--34) as a function of the albedo for the stellar
radiation. We find that the factor decreases from equation (32) to (34),
in other words, as a function of the grain size. The factor 
$\Phi_{\rm i,(3)}^{1/4}$ gives the reduction of the interior
temperature, $T_{\rm i}$, in the three-layer model by isotropic
scattering relative to that in the two-layer model without
scattering. For typical grain sizes found in protoplanetary discs of
0.1--10 \micron, equation (33) would give a good approximation for the
reduction factor. If $\omega_*\approx1$, equations (32) and (33) are
reduced to $\approx\chi_*=\sqrt{1-\omega_*}$. In this case, we expect
the reduction factor of $T_{\rm i}$ to be $\approx(1-\omega_*)^{1/8}$
which is found in Figure~7 later.

\subsubsection{Comparison between numerical and analytic results}

\begin{figure}
 \centering
 \includegraphics[width=7cm]{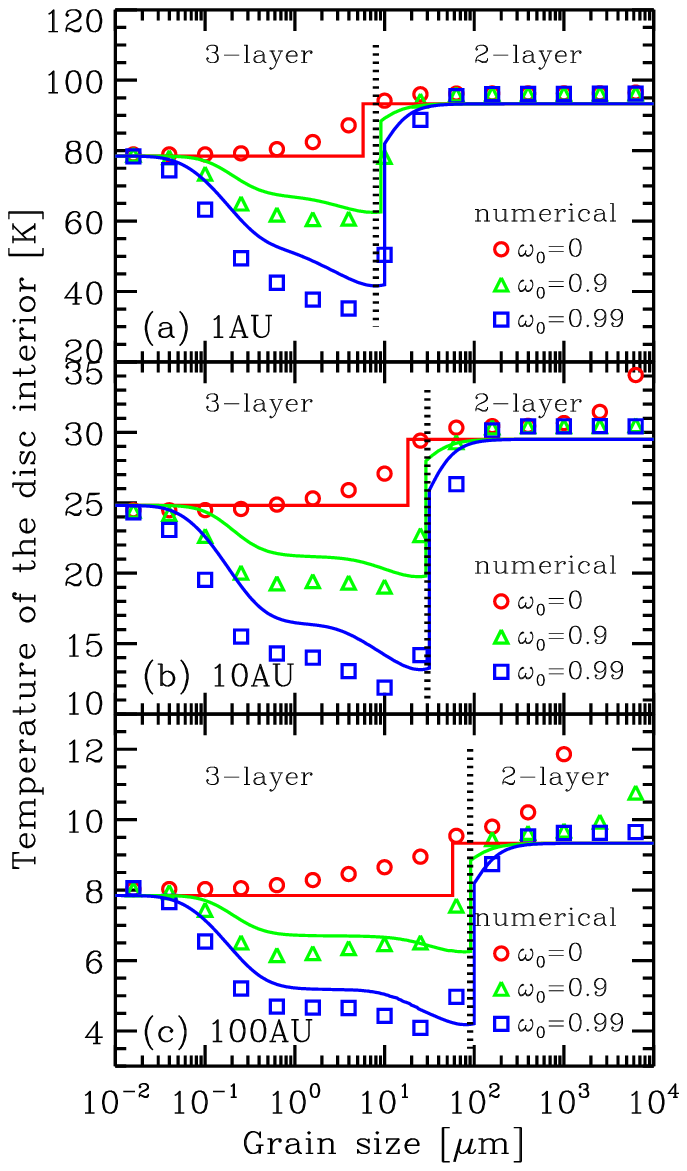}
 \caption{Temperature of the disc interior expected in isotropic
 scattering case as a function of grain size: (a) annulus with the
 radius of 1 AU, (b) 10 AU, and (c) 100 AU. The numerical solutions are
 shown by symbols; circles are no scattering case, triangles are the
 case with the single scattering albedo for small wavelength
 $\omega_0=0.9$, and squares are the case with $\omega_0=0.99$. The
 solid lines are the analytic model described in sections 3.3.1
 (two-layer model; right-hand side) and 3.3.2 (three-layer model;
 left-hand side). We connect these two models around a grain size
 indicated by the dotted line, where a jump appears because of this
 connection (see text in detail). Upwards deviations of numerical
 solutions from the two-layer model found in panels (b) and (c) are
 caused by that the interiors of these cases are optically thin for
 these own radiation.}
\end{figure}

Figure~7 shows the temperature of the interior as a function
of the grain size. The numerical solutions presented in section 2 are
shown by symbols: circles for the no scattering case, triangles for 
the $\omega_0=0.9$ case, and squares for the $\omega_0=0.99$ case. 
The analytic models are shown by the solid lines. 
In analytic models, we always assume the optically thick interior. 
Most of the numerical solutions shown in Figure~7 are
really optically thick. However, when the grain size is larger than
about 100 \micron, because of the reduction of $\kappa_0^{\rm abs}$
given by equation (4), some cases without scattering and with 
$\omega_0=0.9$ become optically thin. In such cases, we find relatively
high temperatures.

The solutions of the analytic models are obtained as follows: 
for the two-layer model, we solved equation (24) to obtain $T_{\rm i}$. 
We need an iterative procedure because the term $\Phi_{\rm i(2)}$
depends on $T_{\rm i}$. For the three-layer model, we obtained 
$T_{\rm i}$ from equation (30) after obtaining $T_{\rm m}$ from 
\begin{equation}
 T_{\rm m} = \left[\alpha W_* B_* 
  \left(\frac{\kappa_{\rm s}^{\rm ext}}{\kappa_{\rm m}^{\rm abs}}\right)
  \Phi_{\rm input}\right]^{1/4}\,,
\end{equation}
which is derived from equations (13), (14), and (26). 
We again need an iterative procedure to obtain both of $T_{\rm m}$ and
$T_{\rm i}$. 

As discussed in \S3.1, the two-layer model is valid only when 
$\kappa_{\rm i}^{\rm ext}/\kappa_{\rm s}^{\rm ext}\approx1$. Otherwise,
we should adopt the three-layer model. Here, we connect these two models
at the grain size where 
$\kappa_{\rm i}^{\rm ext}/\kappa_{\rm s}^{\rm ext}=0.8$ in
the two-layer model, for example; we adopt the two-layer model for a
larger grain size than it and adopt the three-layer model for a smaller
grain size than it. This threshold is rather arbitrary, but we find that
this choice is good as seen in Figure~7 where the connected analytic
models reproduce the numerical solutions reasonably well. Note that a
sudden jump on the solid lines in Figure~7 is caused by this connection
and numerical solutions also show relatively rapid change of the
temperature around there.

When there is no scattering, we find two asymptotic values of $T_{\rm i}$. 
Interestingly, $T_{\rm i}$ for smaller grain size is lower than that for
larger grain size, whereas $T_{\rm s}$ show the opposite trend in
Figure~5. Note that the radiation flux input from the super-heated
layer is in fact independent of $T_{\rm s}$ as shown in equation (20) in
the no scattering case. Nevertheless, the numerical solutions show a
factor of $(1/2)^{1/4}$ reduction of $T_{\rm i}$ for small grain
cases. This is excellently explained by the three-layer model as found
in equation (31). The physical explanation is as follows: when the grain
size is small, the radiation flux from the super-heated layer is
absorbed by the middle layer once. Then, the middle layer emits a half
of the absorbed energy downwards but the rest of the half goes
upwards. Therefore, the interior does not receive a half of the stellar
radiation energy absorbed by the super-heated layer but receive only a
quarter of the energy when the grain size is small.

When there is isotropic scattering, we find further reduction of 
$T_{\rm i}$ in a range of the grain size $a=0.1$--10 or 100 \micron. For 
$a\sim0.01$ \micron, the scattering effect is not observed because of
negligible albedo. For $a\ga10$--100 \micron, the scattering effect is not
observed, either (but except for the optically thin cases). This is nicely
explained by the two-layer model with grey isotropic scattering as found 
in equation (25). The physical reason is that the grey isotropic
scattering equally reduces the downwards flux from the super-heated
layer (eq.~[19]) and the upwards flux from the interior (eq.~[15]). 
The reduction of $T_{\rm i}$ due to scattering for $a=0.1$--10 \micron\ 
is explained by the three-layer model as summarised in equations (32),
(33), and (34). The physical reason of the $T_{\rm i}$ reduction is that
the scattering reduces only the downwards fluxes from the super-heated
layer and from the middle layer because the scattering albedo at the
frequency of the interior radiation is still negligible.

\section{Conclusion}

We have examined the effect of scattering of the diffuse radiation on
the vertical temperature structure of protoplanetary discs. This is
motivated by the fact that scattering albedo increases as the size of
dust grains grows in the discs. In particular, large icy grains have a
significant albedo even in the infrared wavelength. For this aim, 
we have developed a 1D plane-parallel numerical radiation transfer
code including isotropic scattering of the diffuse radiation as well as
that of the incident radiation. We have also developed an analytic model
with isotropic scattering both of the diffuse and the incident
radiations in order to interpret the solutions obtained from the
numerical simulation. All results of the numerical simulations has been
nicely reproduced by the analytic model.

The analytic model presented in this paper is an extension of the
seminal two-layer model by \cite{cg97}; we have introduced a new layer
between the super-heated surface layer and the disc interior. This
middle layer (or disc atmosphere) is required when the absorption of
the radiation from the super-heated layer occurs well above the
photosphere of the opaque isothermal interior. This situation is
realised if the dust opacity is negatively proportional to the
wavelength. Thus, we should consider three layers rather than two layers
if the grain size is smaller than about 10 \micron. On the other hand,
for grey opacity, which is realised if the grain size is $\ga10$
\micron, the standard treatment with two layers is justified.

We have found from the numerical simulation that the dust temperature of
the disc surface is almost not affected by scattering. This is because
the temperature is determined by the radiation equilibrium of grains in
the incident radiation field locally. The grain size has an effect
on the surface temperature via the wavelength dependence of the dust
opacity. There are two asymptotic temperatures: the higher temperature is
realised by small grains ($\la0.1$ \micron) and the lower temperature is
realised by large grains ($\ga1$--10 \micron). This is because the
emission efficiency relative to absorption efficiency of the small
grains is smaller than that of the large grains which have grey
opacity. This trend of the numerical solutions is excellently
reproduced by the analytic model of the super-heated layer.

The numerical simulations without scattering show that the dust
temperature of the optically thick interior also has two asymptotic
values: the lower temperature for small grains ($\la0.1$ \micron) and
the higher temperature for large grains ($\ga10$--100 \micron). Thus,
the trend is opposite from the surface temperature. The higher
asymptotic temperature for large grains is well matched with the
prediction of the two-layer model. On the other hand, the lower
asymptotic temperature for small grains is a factor of $(1/2)^{1/4}$
lower than the prediction of the two-layer model. In fact, the flux
input from the super-heated layer is always same although the interior
temperature is different as a function of the grain size. This phenomenon
has been already found by \cite{dul03a} who attributed it to the energy
loss from the interior by the radiation at a long wavelength 
\citep[see also][]{dul02}. We have proposed another interpretation by
the middle layer between the super-heated layer and the interior; the
super-heated layer gives a half of the absorbed energy of the incident
radiation to the middle layer which gives a half of the obtained energy
to the interior. This three-layer model exactly predicts a factor of
$(1/2)^{1/4}$ reduction of the interior temperature for the small grain
case.

The numerical simulations with isotropic scattering also show two
asymptotic temperatures of the interior. Interestingly, these asymptotic
temperatures of no scattering and isotropic scattering cases are the
same. For small grains ($\la0.1$ \micron), since the scattering albedo
is negligible for all wavelengths interest, the same temperature is
trivial. The same temperature for large grains ($\ga10$--100 \micron)
has been nicely explained by the two-layer model with grey
opacity. The physical mechanism is the exact offset between the
reduction of the flux input into the interior by scattering in the
super-heated layer and the reduction of the flux output from the
interior by scattering in itself (i.e. green-house effect).

For grain sizes of 0.1--10 \micron, which are expected by a moderate
growth of the size in the discs, we have found a further reduction of
the interior temperature in isotropic scattering cases relative to that
without scattering from the numerical simulations. This reduction has
been well explained by the three-layer analytic model. The physical
mechanism is the wavelength dependence of albedo; the flux input into
the interior is reduced by scattering in the super-heated and middle
layers, whereas the flux output from the interior is not reduced because
of negligible (or weak) scattering. Note that the interior flux has a
longer wavelength typically, thus, the albedo for the radiation is
smaller. 

In conclusion, the scattering of the diffuse radiation can affect the
vertical temperature structure of protoplanetary discs significantly
when the grain size grows to be about 1--10 \micron. We need to
investigate the effect in a global disc model in future. The analytic
model presented in this paper could be useful to understand the physics
determining the temperature structure in the discs.

\section*{Acknowledgments}

We would appreciate comments from the anonymous referee which were very
useful for us to improve the quality of this paper.
AKI is grateful to all members of the Department of Physics, Nagoya
University, especially the $\Omega$ Laboratory led by Tsutomu T.\
Takeuchi, for their hospitality during this work. AKI is supported by
KAKENHI (the Grant-in-Aid for Young Scientists B: 19740108) by The
Ministry of Education, Culture, Sports, Science and Technology (MEXT) of
Japan.

\appendix

\section{Variable Eddington factor method with isotropic scattering in
 plane-parallel slab}

We here present our numerical radiation transfer method in a
plane-parallel medium in detail. The method is based on that developed
by \cite{dul02}, but is extended to treat isotropic scatterings of both
of the incident radiation (from the central star) and the diffuse
radiation (from dust grains in the medium). We find a solution, in which
the radiation field, the temperature structure, and the density
structure are consistent with each other, iteratively as the following
procedure:
\begin{enumerate}
 \item assuming an initial temperature structure 
 \item solving the density structure consistent with the given
       temperature structure under the hydrostatic equilibrium
 \item solving the transfer of the incident (stellar) radiation with the
       grazing angle recipe
 \item solving the transfer of the diffuse radiation with a variable
       Eddington factor method and obtaining the temperature structure
       under the radiation equilibrium
 \item checking the convergence of the temperature structure and if not
       going back to the step (ii)
\end{enumerate}
In the following we describe the set of equations and assumptions and
the result of a benchmark test.

\subsection{Hydrostatic equilibrium}

Suppose an annulus clipped from a protoplanetary disk to be a
plane-parallel medium. We set the coordinate $z$ as the vertical height
of the medium. The origin $z=0$ is the equatorial plane of the annulus
and we set a mirror boundary condition there. Assuming the vertical
hydrostatic equilibrium, we obtain the density of gas in the medium
$\rho(z)$ consistent with the temperature structure $T(z)$ which is
assumed as an initial guess or is obtained by the previous step of the
iteration. We assume that the gas temperature is the same as the dust
temperature which is determined by the radiation equilibrium. This
assumption is usually well established in the protoplanetary disc
because the collision between gas particles and dust grains occurs
enough frequently.

The vertical hydrostatic equilibrium is given by 
\begin{equation}
 \frac{dP}{dz} = -\rho g\,,
\end{equation}
where $P$ is the gas pressure and $g$ is the gravitational acceleration.
In a protoplanetary disc, the self-gravity of the disc is negligible
relative to the gravity of the central star. Thus, we have
$g=GM_*z/R^3$, where $G$ is the gravitational constant, $M_*$ is the
mass of the central star, and $R$ is the distance from the star (or
radius of the annulus considered). We have assumed $R\gg z$. The gas
pressure is given by the equation of state for the ideal gas as
$P=(\rho k_{\rm B}T)/(\mu_{\rm m} m_{\rm p})$, where $\mu_{\rm m}$ is the
mean molecular weight, $m_{\rm p}$ is the proton mass, and $k_{\rm B}$
is the Boltzmann constant. Then, equation (A1) is reduced to 
\begin{equation}
 \frac{d\ln \rho}{dz} = -\left(\frac{\mu m_{\rm p}}{k_{\rm B}T}\right)
  \left(\frac{GM_*}{R^3}\right)z + \frac{d\ln T}{dz}\,.
\end{equation}
If we integrate equation (A2) from $z=0$ with a given $T(z)$, we obtain
the functional shape of $\rho(z)$. The absolute value of $\rho(z)$ is
scaled by 
\begin{equation}
 \Sigma = 2\int_0^{z_{\rm max}} \rho(z) dz\,,
\end{equation}
where $\Sigma$ is the gas column density and $z_{\rm max}$ is the
maximum height for the numerical calculation. 
We set $z_{\rm max}=R$ in our calculation and we set the minimum value
of $\rho=10^{-25}$ g cm$^{-3}$ just for avoiding a too small value of
the density.

We set the optical depth coordinate $\tau$ for the radiation transfer
from the obtained $\rho(z)$ as 
\begin{equation}
 \tau_\nu(z) = \int_z^{z_{\rm max}} \rho(z) \kappa_\nu^{\rm ext} dz\,,
\end{equation}
where $\kappa_\nu^{\rm ext}$ is the extinction cross section by dust
grains per unit gas mass at the frequency $\nu$.

In fact, we can obtain the temperature structure as a function of the
optical depth $T(\tau)$ by the radiation transfer without the density
structure as a function of the vertical height $\rho(z)$. The reason
why we calculate $\rho(z)$ and the relation between the optical depth
and the vertical height $\tau(z)$ is to see $T(z)$ as shown in Figure~3.
Another reason, which may be more important, is to determine the grazing
angle consistent with the disc global structure for future calculations.

\subsection{Transfer of the incident radiation}

Let us consider an incident radiation beam entering the plane-parallel
medium. The angle between the incident ray and the surface of
the medium is called the grazing angle, $\alpha$. In a protoplanetary
disc, the grazing angle $\alpha$ is usually as small as $\sim0.05$
radian \citep[e.g.,][]{dal06}. Thus, we use the approximation 
$\sin\alpha\approx\alpha$. The optical depth along the incident ray
becomes $\tau_\nu(z)/\alpha$. Thus, the mean intensity of
the (direct) incident radiation is given by 
\begin{equation}
 J^*_\nu(z) = J^{*{\rm max}}_\nu e^{-\tau_\nu(z)/\alpha}\,,
\end{equation}
where $J^{*{\rm max}}_\nu$ is the mean intensity at the top of the
medium (i.e. $z=z_{\rm max}$). For the incident radiation from the
central star, $J^{*{\rm max}}_\nu=B_\nu(T_*) \Omega_*/4\pi$, where 
$B_\nu(T_*)$ is the Planck function with the stellar effective
temperature $T_*$ and $\Omega_*=\pi(R_*/R)^2 f_{\rm vis}$ is the solid
angle of the stellar photosphere visible from the top of the medium 
($f_{\rm vis}$ is the visible fraction of the stellar photosphere).
Finally, we give the absorbed and extincted energy density of the
incident radiation per unit time interval at the height $z$ as 
\begin{equation}
 q_{\rm abs}(z)=\int_0^\infty \rho(z)\kappa_\nu^{\rm abs}4\pi J_\nu^* d\nu\,,
\end{equation}
and 
\begin{equation}
 q_{\rm ext}(z)=\int_0^\infty \rho(z)\kappa_\nu^{\rm ext}4\pi J_\nu^* d\nu\,,
\end{equation}
where $\kappa_\nu^{\rm abs}$ and $\kappa_\nu^{\rm ext}$ are the
absorption and the extinction cross section by dust per unit gas mass at
the frequency $\nu$.

\subsection{Transfer of the diffuse radiation}

The transfer equation of the diffuse radiation (or the radiation
reprocessed by dust) in a plane-parallel medium is 
\begin{equation}
 \mu\frac{dI_{\nu\mu}}{dz} = 
  -\rho\kappa_\nu^{\rm ext}I_{\nu\mu}
  +\rho\kappa_\nu^{\rm ext}S_\nu\,,
\end{equation}
where $\mu$ is the cosine of the angle between the ray and the $z$
coordinate, $I_{\nu\mu}$ is the specific intensity at the frequency
$\nu$ towards the direction $\mu$, and $S_\nu$ is the source function
which is given by  
\begin{equation}
 S_\nu = (1-\omega_\nu) B_\nu(T) + \omega_\nu J_\nu + \omega_\nu J_\nu^*\,,
\end{equation}
where $\omega_\nu$ is the single scattering albedo at the frequency
$\nu$, $B_\nu(T)$ is the Planck function with the dust temperature $T$, 
and $J_\nu$ is the mean intensity of $I_{\nu\mu}$, that is, 
$\displaystyle J_\nu=\frac{1}{2}\int_{-1}^1 I_{\nu\mu}d\mu$.
We have assumed that the scattering is isotropic for the
simplicity. Note that we consider the scattering of the diffuse
radiation as the second term in equation (A9). The third term accounts
for the scattering of the incident radiation. To determine the dust
temperature $T$, we assume the radiation equilibrium as 
\begin{equation}
 \int_0^\infty \rho \kappa_\nu^{\rm abs} B_\nu(T) d\nu
  = \int_0^\infty \rho \kappa_\nu^{\rm abs} J_\nu d\nu 
  + \frac{q_{\rm abs}}{4\pi}\,.
\end{equation}

To obtain the mean intensity $J_\nu$,
we adopt a variable Eddington factor method. The first and second
moments of equation (A8) are 
\begin{equation}
 \frac{dH_\nu}{dz} = \rho \kappa_\nu^{\rm abs} (B_\nu - J_\nu)
  + \rho \kappa_\nu^{\rm sca} J_\nu^*\,,
\end{equation}
and 
\begin{equation}
 \frac{dK_\nu}{dz} = - \rho \kappa_\nu^{\rm ext} H_\nu\,,
\end{equation}
where $\displaystyle H_\nu=\frac{1}{2}\int_{-1}^1I_{\nu\mu}\mu d\mu$ and 
$\displaystyle K_\nu=\frac{1}{2}\int_{-1}^1I_{\nu\mu}\mu^2d\mu$, and we
have used equation (A9) to eliminate $S_\nu$. Note that 
$\kappa_\nu^{\rm abs}=(1-\omega_\nu)\kappa_\nu^{\rm ext}$ and 
the scattering cross section per unit gas mass 
$\kappa_\nu^{\rm sca}=\omega_\nu \kappa_\nu^{\rm ext}$. If we integrate
equations (A11) and (A12) over the frequency, we obtain 
\begin{equation}
 \frac{dH}{dz} = \frac{q_{\rm ext}}{4\pi}\,,
\end{equation}
and 
\begin{equation}
 \frac{dK}{dz} = -\int_0^\infty \rho \kappa_\nu^{\rm ext} H_\nu d\nu\,,
\end{equation}
where $\displaystyle H=\int_0^\infty H_\nu d\nu$ and 
$\displaystyle K=\int_0^\infty K_\nu d\nu$, and we have used the
radiation equilibrium (eq.[A10]) and equations (A6) and (A7) in equation 
(A13). Finally, we introduce the Eddington factor as the closure equation: 
\begin{equation}
 f_{\rm E} = \frac{K}{J}\,,
\end{equation}
where $\displaystyle J=\int_0^\infty J_\nu d\nu$. 

We do not assume $f_{\rm E}=1/3$ (constant) as in the usual Eddington
approximation, but obtain $f_{\rm E}$ directly from $I_{\nu\mu}$ which
is calculated by the formal solution of equation (A8). The integration
of the formal solution is performed with a parabolic interpolation of
$S_\nu$ among three successive spatial points \citep{ols87}. As described
in \cite{dul02}, we alternate the integration of the formal solution
(i.e. ray-tracing) with the integration of the moment equations
(eqs.~[A13]--[A15]) and the determination of $T$ by equation~[A10] until
we reach a convergence in $T$ (the difference between the two successive
iterations becomes less than 0.1\%). In addition, we adopt the
acceleration algorithm by \cite{ng74} for a rapid convergence. For no
scattering case, the convergence is very rapid independent of the total
optical depth of the medium, typically 20 iterations. On the other hand,
a factor of $\sim10$ times more iterations depending on the albedo are
needed for isotropic scattering case.

\subsection{Benchmark test}

\begin{figure}
 \centering
 \includegraphics[width=7cm]{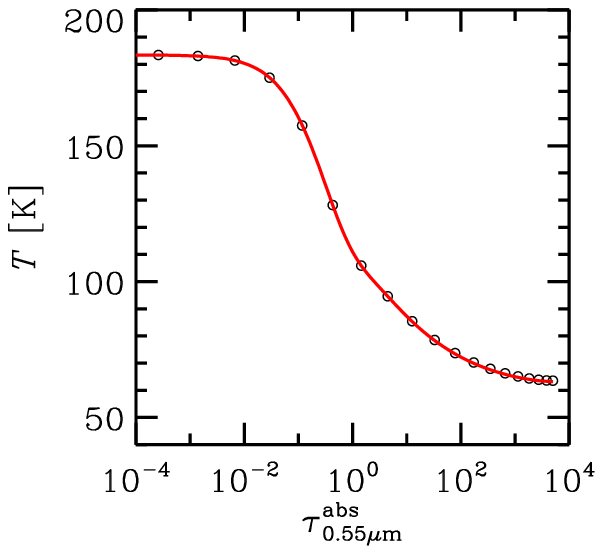}
 \caption{Result of a benchmark test without scattering proposed by
 C.~P.~Dullemond. The circles are the reference solutions by
 C.~P.~Dullemond, whereas the solid line is our solution.}
\end{figure}

Figure~A1 shows the result of the benchmark test No.3 proposed by
C.~P.~Dullemond on his web page
\footnote{http://www.mpia-hd.mpg.de/homes/dullemon/radtrans/benchmarks/}.
The settings are as follows: the stellar temperature $T_*=3,000$ K, 
the stellar radius $R_*=2.0$ $R_\odot$, the distance from the star $R=1$
AU, the grazing angle $\alpha=0.05$, the visible fraction of the stellar
photosphere $f_{\rm vis}=0.5$, the total visual optical depth of the
disc $\tau_{0.55\micron}=10^4$ (i.e. $\tau_{0.55\micron}=5\times10^3$ at
the equatorial plane), and no scattering. We used the dust opacity model
downloaded from the web page which is the same as that assumed in the
calculation by C.~P.~Dullemond. The figure shows an excellent agreement
between both results.

\section{Analytic expression of radiation field in isothermal,
 absorption, and isotropic scattering medium}

Here, we derive an analytic expression of the radiation field in an
isothermal medium with absorption and isotropic scattering. 
In the derivation, we adopt the Eddington approximation with two-stream
lines \citep[e.g.,][]{ryb79}. We consider a diffuse incident radiation.

\begin{figure}
 \centering
 \includegraphics[width=7cm]{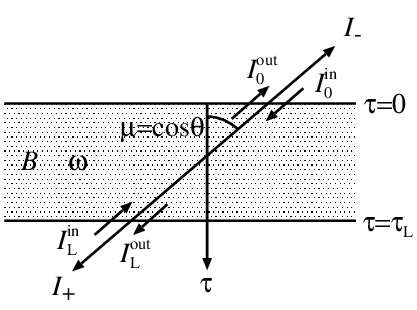}
 \caption{Plane-parallel isothermal medium with isotropic scattering.}
\end{figure}

\begin{figure}
 \centering
 \includegraphics[width=7cm]{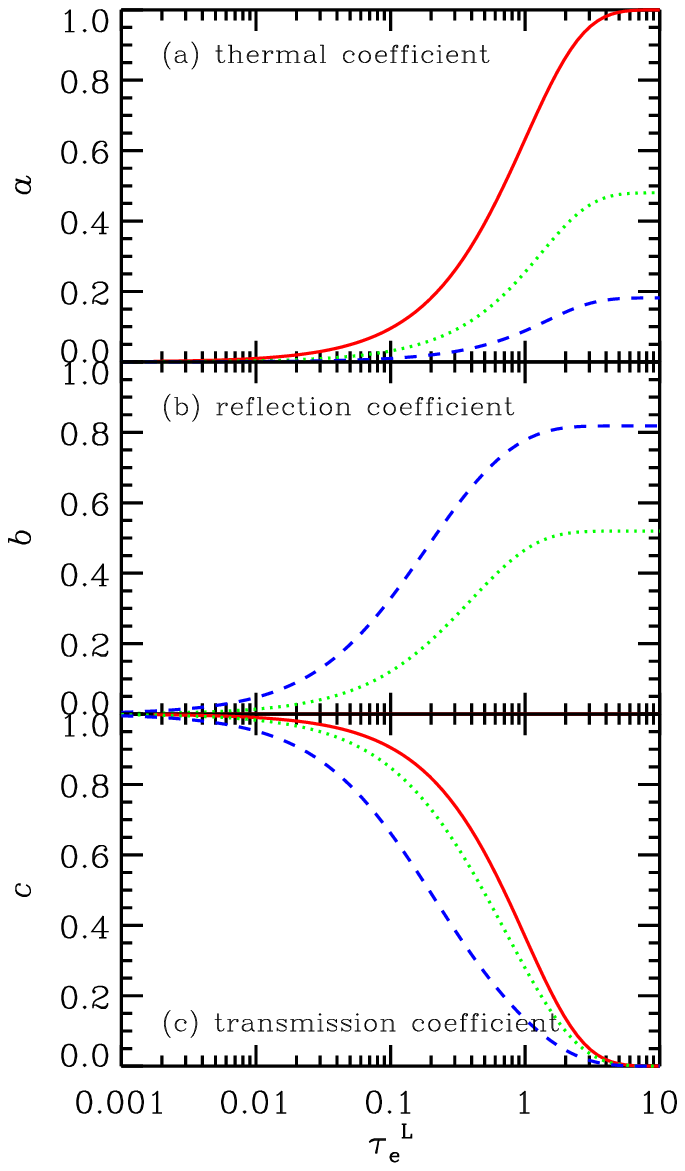}
 \caption{Thermal, reflection, and transmission coefficients as a
 function of effective optical depth of an isothermal, absorption, and
 isotropic scattering medium. In each panel, we show three cases of the
 single scattering albedo: $\omega=0$ (solid curve), $0.90$ (dotted
 curve), and $0.99$ (dashed curve). The reflection coefficient $b$ is
 always zero when $\omega=0$, so that we cannot see the solid curve in
 the panel (b). The effective optical depth 
 $\tau_{\rm e}^{\rm L}=\sqrt{3(1-\omega)}\,\tau_{\rm L}$ if the total
 (absorption+scattering) optical depth of the medium is $\tau_{\rm L}$. 
 Asymptotic values of the coefficients are given in equations 
 (B17)--(B19).}
\end{figure}

Suppose a plane-parallel medium (see Figure~B1). We set the extinction
(absorption+scattering) optical depth coordinate $\tau$ along the normal
of the medium. The total extinction optical depth of the medium is set
to be $\tau_{\rm L}$. Then, suppose that the single scattering albedo in
this medium is $\omega$ and this medium is isothermal and in the thermal
equilibrium. The thermal radiation is denoted as $B$.
Let us consider two-stream lines with the direction of 
$\mu=\pm1/\sqrt{3}$, where $\mu$ is the cosine of the angle between the
ray and the $\tau$ coordinate. When the direction of the rays is denoted
by the subscript of $+$ or $-$, the equation of the radiation transfer
becomes 
\begin{equation}
 \pm\frac{1}{\sqrt{3}} \frac{dI_{\pm}}{d\tau} = S_{\pm} - I_{\pm}\,.
\end{equation}
The source function can be expressed as 
\begin{equation}
 S_{\pm}=(1-\omega)B + \omega J\,,
\end{equation}
where $J$ is the mean intensity.
In the two-stream approximation, we can define the mean intensity $J$,
the mean flux $H$, and the mean radiation pressure $K$ as  
\begin{equation}
 J = \frac{1}{2}(I_+ + I_-)\,,
\end{equation}
\begin{equation}
 H = \frac{1}{2\sqrt{3}}(I_+ - I_-)\,,
\end{equation}
and 
\begin{equation}
 K = \frac{1}{6}(I_+ + I_-) = \frac{1}{3}J\,.
\end{equation}

The first and second moments of (B1) with the source function (B2) are 
\begin{equation}
 \frac{dH}{d\tau} = (1-\omega)(B-J)\,,
\end{equation}
and 
\begin{equation}
 \frac{1}{3}\frac{dJ}{d\tau} = - H\,.
\end{equation}
With an effective optical depth 
$\tau_{\rm e}\equiv \tau\sqrt{3(1-\omega)}$, we have 
\begin{equation}
 \frac{d^2 J}{d\tau_{\rm e}^2} = J - B\,, 
\end{equation}
from equations (B6) and (B7).

If we have incident radiations at the upper and lower boundaries as 
$I_0^{\rm in}$ and $I_{\rm L}^{\rm in}$, respectively, the boundary
conditions are 
\begin{equation}
 J(0)=I_0^{\rm in}+\sqrt{1-\omega}
  \left(\frac{dJ}{d\tau_{\rm e}}\right)_{0}\,,
\end{equation}
and 
\begin{equation}
 J(\tau_{\rm e}^{\rm L}) = 
  I_{\rm L}^{\rm in}-\sqrt{1-\omega}
  \left(\frac{dJ}{d\tau_{\rm e}}\right)_{\rm L}\,,
\end{equation}
where $\tau_{\rm e}^{\rm L}=\tau_{\rm L}\sqrt{3(1-\omega)}$.
The solution of equation (B8) with the boundary conditions (B9) and (B10)
is 
\begin{eqnarray}
 J(\tau_{\rm e}) & = & B
  \left[1-\frac{e^{-\tau_{\rm e}}+e^{-\tau_{\rm e}^{\rm L}+\tau_{\rm e}}}
   {1+\chi+(1-\chi)e^{-\tau_{\rm e}^{\rm L}}}\right] 
  \cr
  & + & I_0^{\rm in}
  \frac{(1+\chi)e^{-\tau_{\rm e}}
  -(1-\chi)e^{-2\tau_{\rm e}^{\rm L}+\tau_{\rm e}}}
  {(1+\chi)^2 - (1-\chi)^2 e^{-2\tau_{\rm e}^{\rm L}}}
  \cr
  & + & I_{\rm L}^{\rm in}
  \frac{(1+\chi)e^{-\tau_{\rm e}^{\rm L}+\tau_{\rm e}}
  -(1-\chi)e^{-\tau_{\rm e}^{\rm L}-\tau_{\rm e}}}
  {(1+\chi)^2 - (1-\chi)^2 e^{-2\tau_{\rm e}^{\rm L}}}
  \,,
\end{eqnarray}
where $\chi=\sqrt{1-\omega}$. In the case without incident radiation 
(i.e. $I_0^{\rm in}=I_{\rm L}^{\rm in}=0$), this solution is
exactly same as equation~(28) of \cite{miy93}.

Then, let us consider the outbound intensity at the surface of the
medium. At $\tau=0$, the outbound intensity can be 
\begin{equation}
 I_{0}^{\rm out} = a B + b I_{0}^{\rm in} + c I_{\rm L}^{\rm in}\,,
\end{equation}
where $a$, $b$, and $c$ can be called as ``thermal'', ``reflection'',
and ``transmission'' coefficients, respectively. Since
$J(0)=(1/2)(I_0^{\rm in}+I_0^{\rm out})$, we obtain from equation (B11) 
\begin{equation}
 a = \frac{2\chi(1-e^{-\tau_{\rm e}^{\rm L}})}
  {1+\chi + (1-\chi)e^{-\tau_{\rm e}^{\rm L}}}\,,
\end{equation}
\begin{equation}
 b = \frac{(1-\chi)(1+\chi)
  (1-e^{-2\tau_{\rm e}^{\rm L}})}
  {(1+\chi)^2 - (1-\chi)^2 
  e^{-2\tau_{\rm e}^{\rm L}}}\,,
\end{equation}
and 
\begin{equation}
 c = \frac{4\chi\,e^{-\tau_{\rm e}^{\rm L}}}
  {(1+\chi)^2 - (1-\chi)^2 
  e^{-2\tau_{\rm e}^{\rm L}}}\,.
\end{equation}
At $\tau=\tau_{\rm L}$, we obtain symmetrically 
\begin{equation}
 I_{\rm L}^{\rm out} = a B + b I_{\rm L}^{\rm in} + c I_{0}^{\rm in}\,.
\end{equation}

Figure~B2 shows the three coefficients for an isotropic case as a
function of the effective optical depth of the medium
$\tau_{\rm e}^{\rm L}$. As limiting values, we obtain 
\begin{equation}
 a \to \cases{
  \displaystyle\frac{2\chi}{1+\chi} 
  & ($\tau_{\rm L}\to\infty$) \cr
  0 & ($\tau_{\rm L}\to 0$) \cr
  }\,,
\end{equation}
\begin{equation}
 b \to \cases{
  \displaystyle\frac{1-\chi}{1+\chi} 
  & ($\tau_{\rm L}\to\infty$) \cr
  0 & ($\tau_{\rm L}\to 0$) \cr
  }\,,
\end{equation}
and 
\begin{equation}
 c \to \cases{
  0 & ($\tau_{\rm L}\to\infty$) \cr
  1 & ($\tau_{\rm L}\to 0$) \cr
  }\,.
\end{equation}

\label{lastpage}


\begin{thebibliography}{99}


\bibitem[Calvet et al.(1991)]{cal91}
Calvet, N., Patino, A., Magris, G. C., D'Alessio, P., 
1991, ApJ, 380, 617

\bibitem[Chiang \& Goldreich(1997)]{cg97}
Chiang, E. I., Goldreich, P., 1997, ApJ, 490, 368

\bibitem[Chiang et al.(2001)]{chi01}
Chiang, E. I., Joung, M. K., Creech-Eakman, M. J., Qi, C., 
Kessler, J. E., Blake, G. A., van Dishoeck, E. F., 2001, ApJ, 547, 1077

\bibitem[D'Alessio et al.(2006)]{dal06}
D'Alessio, P., Calvet, N., Hartmann, L., Franco-Hern{\'a}ndez, R., 
Serv{\'i}n, H., 2006, ApJ, 638, 314

\bibitem[Draine(2003)]{dra03}
Draine, B. T., 2003, ARA\&A, 41, 241

\bibitem[Draine \& Lee(1984)]{dra84}
Draine, B. T., Lee, H.-M., 1984, ApJ, 285, 89

\bibitem[Dullemond \& Turolla(2000)]{dul00}
Dullemond, C. P., Turolla, R., 2000, A\&A, 360, 1187

\bibitem[Dullemond, Dominik, \& Natta(2001)]{dul01}
Dullemond, C. P., Dominik, C., Natta, A., 2001, ApJ, 560, 957

\bibitem[Dullemond et al.(2002)]{dul02}
Dullemond, C. P., van Zadelhoff, G. J., Natta, A., 
2002, A\&A, 389, 464

\bibitem[Dullemond \& Natta(2003a)]{dul03a}
Dullemond, C. P., Natta, A., 2003a, A\&A, 405, 597

\bibitem[Dullemond \& Natta(2003b)]{dul03}
Dullemond, C. P., Natta, A., 2003b, A\&A, 408, 161

\bibitem[Garaud \& Lin(2007)]{gar07}
Garaud, P., Lin, D. N. C., 2007, ApJ, 654, 606

\bibitem[Miyake \& Nakagawa(1993)]{miy93}
Miyake, K., Nakagawa, Y., 1993, Icarus, 106, 20

\bibitem[Ng(1974)]{ng74}
Ng, K., 1974, J. Chem. Phys., 61, 2680

\bibitem[Nomura et al.(2008)]{nom08}
Nomura, H., Aikawa, Y., Nakagawa, Y., Millar, T. J., 
2008, A\&A, in press (arXiv:0810.4610)

\bibitem[Oka et al.(2009)]{oka09}
Oka, A., Nakamoto, T., et al.
2009, in preparation

\bibitem[Olson \& Kunasz(1987)]{ols87}
Olson, G. L., Kunasz, P. B., 1987, JQSRT, 38, 325

\bibitem[Pascucci et al.(2004)]{pas04}
Pascucci, I., Wolf, S., Steinacker, J., Dullemond, C. P., Henning, Th., 
Niccolini, G., Woitke, P., Lopez, B., 2004, A\&A, 417, 793

\bibitem[Rafikov \& De Colle(2006)]{raf06}
Rafikov, R. R., De Colle, F., 2006, ApJ, 646, 275

\bibitem[Rybicki \& Lightman(1979)]{ryb79}
Rybicki, G. B., Lightman, A. P., 
1979, Radiative Processes in Astrophysics, 
Wiley-Interscience, New York

\bibitem[Sano et al.(2000)]{san00}
Sano, T., Miyama, S., Umebayashi, T., Nakano, T., 
2000, ApJ, 543, 486

\bibitem[Sasselov \& Lecar(2000)]{sas00}
Sasselov, D. D., Lecar, M., 2000, ApJ, 528, 995

\bibitem[Steinacker, Bacmann, \& Henning (2006)]{ste06}
Steinacker, J., Bacmann, A., Henning, Th., 2006, ApJ, 645, 920

\bibitem[Strittmatter(1974)]{str74}
Strittmatter, P. A., 1974, A\&A, 32, 7

\end{thebibliography}
\end{document}